\documentclass[apj,twocolumn]{emulateapj}
\usepackage{amsmath}
\eqsecnum

\renewcommand\email\texttt

\newcommand\nameobj{Leo T }
\newcommand\msun{M_\odot}

\newcommand\coords{{09:34:53.4\, +17:03:05}}
\def\spose#1{\hbox to 0pt{#1\hss}}
\def\lta{\mathrel{\spose{\lower 3pt\hbox{$\sim$}}
    \raise 2.0pt\hbox{$<$}}}
\def\gta{\mathrel{\spose{\lower 3pt\hbox{$\sim$}}
    \raise 2.0pt\hbox{$>$}}}

\begin{document} 

\slugcomment{\sc submitted to \it Astrophysical Journal Letters}
\shorttitle{\sc \nameobj Dwarf} 
\shortauthors{Irwin et al.}

\title{Discovery of an Unusual Dwarf Galaxy in the Outskirts of the
Milky Way}
\author{M.\ J.\ Irwin\altaffilmark{1},
V.\ Belokurov\altaffilmark{1},
N.\ W.\ Evans\altaffilmark{1},
E.\ V.\ Ryan-Weber\altaffilmark{1},
J.\ T.\ A.\ de Jong\altaffilmark{2},
S.\ Koposov\altaffilmark{2},
D.\ B.\ Zucker\altaffilmark{1}, 
S.\ T.\ Hodgkin\altaffilmark{1},
G.\ Gilmore\altaffilmark{1},
P.\ Prema\altaffilmark{1},
L.\ Hebb\altaffilmark{3},
A.\ Begum\altaffilmark{1},
M.\ Fellhauer\altaffilmark{1},
P.\ C.\ Hewett\altaffilmark{1},
R.\ C. Kennicutt, Jr.\altaffilmark{1},
M.\ I.\ Wilkinson\altaffilmark{1},
D.\ M.\ Bramich\altaffilmark{1},
S.\ Vidrih\altaffilmark{1}, 
H.-W.\ Rix\altaffilmark{2},
T.\ C.\ Beers\altaffilmark{4},
J.\ C.\ Barentine\altaffilmark{5},
H.\ Brewington\altaffilmark{5},
M.\ Harvanek\altaffilmark{5},
J.\ Krzesinski\altaffilmark{5,6},
D.\ Long\altaffilmark{5},
A.\ Nitta\altaffilmark{7},
S.\ A.\ Snedden\altaffilmark{5}
}

\altaffiltext{1}{Institute of Astronomy, University of Cambridge,
Madingley Road, Cambridge CB3 0HA, UK;\email{mike,vasily,nwe@ast.cam.ac.uk}}
\altaffiltext{2}{Max Planck Institute for Astronomy, K\"{o}nigstuhl
17, 69117 Heidelberg, Germany}
\altaffiltext{3}{School of Physics and Astronomy, University of St
Andrews, North Haugh, St Andrews KY16 9SS}
\altaffiltext{4}{Department of Physics and Astronomy, Michigan State 
University, East Lansing, MI 48824}
\altaffiltext{5}{Apache Point Observatory, P.O. Box 59, Sunspot, NM 88349}
\altaffiltext{6}{Mt. Suhora Observatory, Cracow Pedagogical
  University, ul. Podchorazych 2, 30-084 Cracow, Poland}
\altaffiltext{7}{Gemini Observatory, 670 N. A'ohoku Place, Hilo, HI 96720}

\begin{abstract}
In this Letter, we announce the discovery of a new dwarf galaxy,
\nameobj, in the Local Group. It was found as a stellar overdensity in
the Sloan Digital Sky Survey Data Release 5 (SDSS DR5). The
color-magnitude diagram of \nameobj shows two well-defined features,
which we interpret as a red giant branch and a sequence of young,
massive stars. As judged from fits to the color-magnitude diagram, it
lies at a distance of $\sim 420$ kpc and has an intermediate-age
stellar population with a metallicity of [Fe/H]= -1.6, together with a
young population of blue stars of age $\sim 200$ Myr. There is a
compact cloud of neutral hydrogen with mass $\sim 10^5 \msun$ and
radial velocity $= +35$ kms${}^{-1}$ coincident with the object
visible in the HIPASS channel maps. \nameobj is the smallest, lowest
luminosity galaxy found to date with recent star-formation. It
appears to be a transition object similar to, but much lower
luminosity than, the Phoenix dwarf.
\end{abstract}


\section{Introduction}

The last two years have seen the discovery of 10 faint, new Milky Way
satellites in data from the Sloan Digital Sky Survey (SDSS). This is
made up of 8 new Milky Way dwarf galaxies, together with 2 unusually
extended globular clusters~\citep{Wi05,Zu06a,Zu06b,Be06a,Be07}. The
purpose of this {\it Letter} is to announce the discovery of an
additional dwarf galaxy in the Local Group. At a heliocentric distance
of $\sim 420$ kpc, this is at the very outskirts of the Milky Way's
sphere of influence.

The dwarf is faint with roughly circular isopleths from the majority
intermediate-age stellar population. These properties are shared by
other Local Group dwarf spheroidal (dSph) galaxies. However, \nameobj
also has a population of bright, blue stars, which is evidence for an
epoch of recent star-formation. It must therefore have contained some
gas in the recent past, and there is evidence of a compact, neutral
gas cloud still associated with the object today. These properties are
characteristic of the dwarf irregular (dIrrs) galaxies.  As the dwarf
possesses transitional properties -- intermediate between those of
dSphs and dIrrs -- we propose to call it \nameobj. Other transitional
dwarf galaxies are known, in particular Phoenix, Pisces and Leo A
~\citep[see e.g.,][]{Gr01}, which are at comparable distances from the
Galaxy and M31.

\begin{figure*}[t]
\begin{center}
\includegraphics[width=0.9\textwidth]{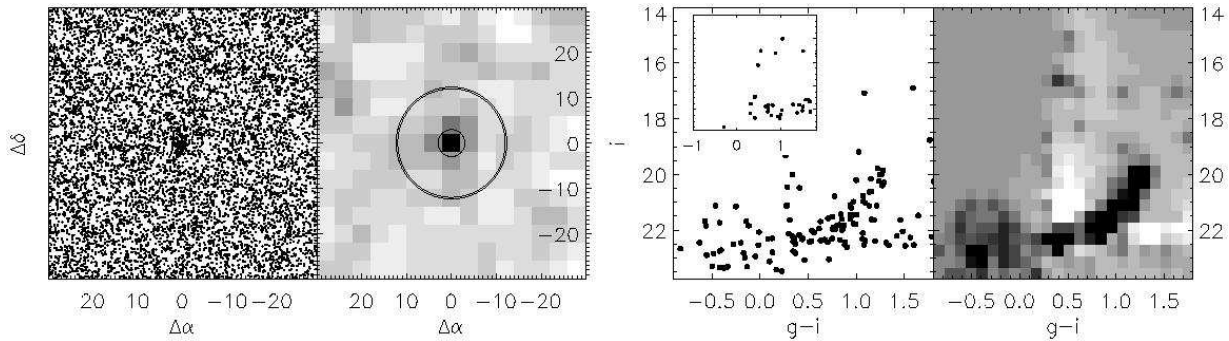}
\caption{The \nameobj Dwarf: {\it Left:} The spatial distribution of
all objects classified as stars in the vicinity of \nameobj. $\Delta
\alpha$ and $\Delta \delta$ are given in arcminutes. {\it Middle
Left:} Binned spatial density of all stellar objects. The inner circle
marks a radius of $3'$, and the outer circle a radius of $12'$. Bins
are $4\farcm \times 4\farcm$, smoothed with a Gaussian with a FWHM of
$6\farcm$ {\it Middle right:} CMD of all stellar objects within the
inner $3'$ radius circle; note the red giant branch (RGB). The inset
shows a control CMD with all stellar objects in an annulus between
$12\farcm0$ and $12\farcm4$ of the center. {\it Right:} A
color-magnitude density plot (Hess diagram), showing the inner CMD
minus a larger area control CMD, appropriately normalised.  The red
giant branch is clearly seen, along with another blue feature
discussed in the main text.
\label{fig:leo_disc}}
\end{center}
\end{figure*}
\begin{deluxetable}{lc}
\tablecaption{Properties of the \nameobj Dwarf \label{tbl:pars}}
\tablewidth{0pt} \tablehead{ \colhead{Parameter\tablenotemark{a}} &
{~~~ } } \startdata Coordinates (J2000) & \coords \\
Coordinates (Galactic) & $\ell = 214.9^\circ, b = 43.7^\circ$ \\
 Position Angle & $\approx 0^{\circ}$\\
 Ellipticity & $\approx 0$\\
 $r_h$ (Plummer) & $1\farcm4$\\
 A$_{\rm V}$ & $0\fm1$ \\
 $\mu_{\rm 0,V}$ (Plummer) & $26\fm9$\\
 $V_{\rm tot}$ & $16\fm0$\\
(m$-$M)$_0$ & $23\fm1$\\
 M$_{\rm tot,V}$ & $-7\fm1$ \\
 $M_{\rm HI}$ & $10^5 \msun$ \\
 ${\rm v}_\odot$\tablenotemark{b} & +35 kms${}^{-1}$ \\
\enddata
\tablenotetext{a}{Surface brightnesses and integrated magnitudes 
are accurate to $\sim \pm 0\fm3$ and are corrected for the 
mean Galactic foreground reddening, A$_{\rm V}$, shown.}
\tablenotetext{b}{From HI data, no stellar velocities measured to date}
\label{tab:struct}
\end{deluxetable}

\section{Data and Discovery}

SDSS imaging data are produced in five photometric bands, namely $u$,
$g$, $r$, $i$, and $z$~\citep{Fu96,Gu98,Ho01,Am06,Gu06}. The data are
automatically processed through pipelines to measure photometric and
astrometric properties \citep{Lu99,St02,Sm02,Pi03,Iv04} and
de-reddened with the help of \citet{Sc98}. Data Release 5 (DR5)
primarily covers $\sim 8000$ square degrees around the North Galactic
Pole (NGP).

As part of our systematic survey of stellar overdensities in DR5
\citep{Be06b,Be07}, we identified a new candidate in the constellation
of Leo. A roughly spherical overdensity of objects classified by the
SDSS pipeline as stars is readily visible in the left panel of
Figure~\ref{fig:leo_disc}. Plotting these stars in a color-magnitude
diagram (CMD) and using it to construct a Hess diagram (right panels
of Figure~\ref{fig:leo_disc}) reveals a clear red giant branch
(RGB). Although sparsely populated, the CMD looks like those of the
dwarf irregular or transition galaxies with detected young, blue
populations, such as Phoenix~\citep{Ma99} and Leo A~\citep{To98}. The
bright, blue sequence in the CMD may be a nearly zero-age main
sequence or blue loop, or even blue straggler stars. Some of
the stars may even be blue horizontal branch stars scattered into this
region of the CMD by photometric errors.  Precise interpretation needs
deeper photometry, but the general conclusion is clear -- the \nameobj
dwarf contains a population of young, recently formed ($<$ 1 Gyr)
stars.

\begin{figure}[t]
\begin{center}
\includegraphics[width=0.4\textwidth]{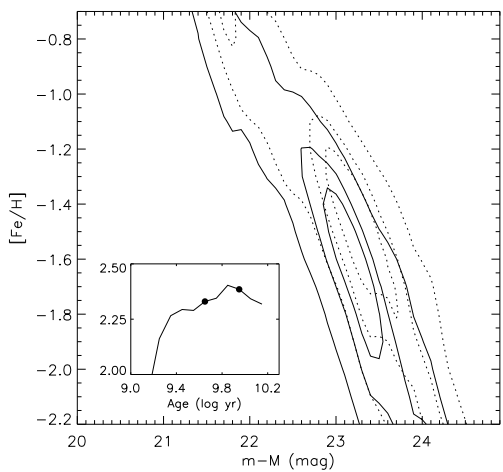}
\caption{Goodness-of-fit contours of CMD fits using the method of de
Jong et al. (2007) for two different populations with ages 4-5 Gyr
(dotted) and 8-10 Gyr (solid). The levels correspond to $0.5, 1$ and
$2$ sigma uncertainties. The inset shows the improvement in the
maximum-likelihood goodness-of-fit parameter Q (Dolphin 2002) with
circles representing the two populations.\label{fig:jelte}}
\end{center}
\end{figure}
\begin{figure}[th]
\begin{center}
\includegraphics[width=0.4\textwidth]{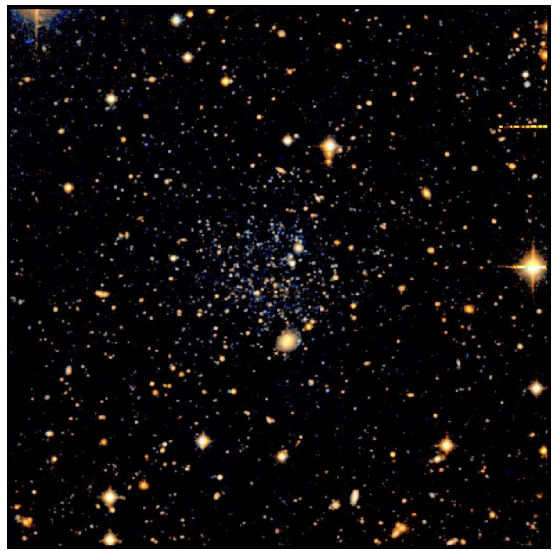}
\caption{Color image covering a $6\farcm5 \times 6\farcm5$ region centered 
on the \nameobj dwarf made from INT WFC data in the $g$ and $r$ bands.
N is to the top and E to the left.
\label{fig:leo_image}}
\end{center}
\end{figure}
\begin{figure*}
\begin{center}
\includegraphics[width=\textwidth]{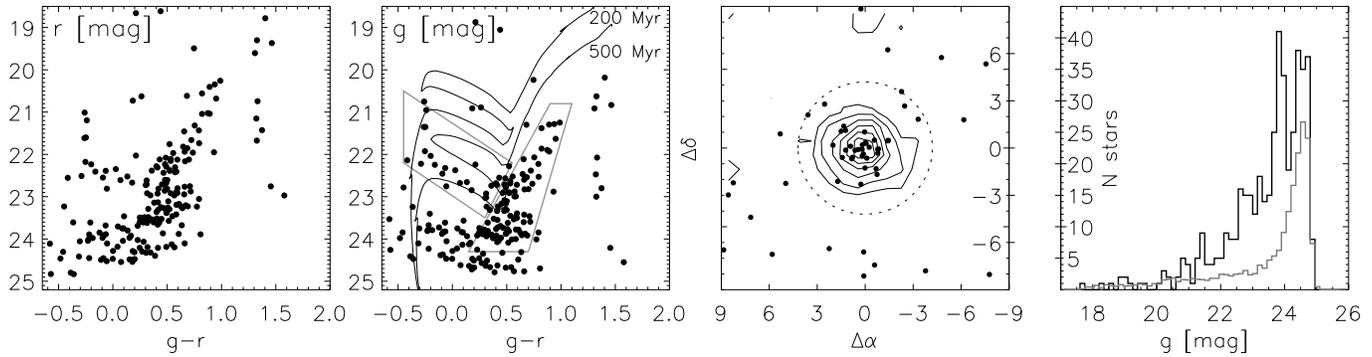}
\caption{Left and middle left: CMDs in $g-r$ versus $r$ and $g$ based
on INT imaging data within a radius of $1\farcm8$ of the center of
\nameobj.  The light grey boxes enclose the RGB and young blue
candidate stars. Overplotted are isochrones from \citet{Gi04} for
[Fe/H] $= -1.7$ and age 200 and 500 Myrs. Middle right: The density
contours of the RGB candidates in a box $18'\times 18'$ centered on
\nameobj, together with the locations of the possible young blue
stars. Right: The $g$ band luminosity function within the dotted
circle (black) compared to the background (grey). There is a clear
detection of the horizontal branch at $g \sim 23.8$.
\label{fig:leo_int}}
\end{center}
\end{figure*}
The distance, metallicity and age of the object were constrained by
applying a colour-magnitude diagram (CMD) fitting technique to the
SDSS photometry. Using the MATCH software~\citep{Do02,deJ07} together
with isochrones from \citet{Gi04}, model stellar populations with
fixed age range, metallicity and distance were fit to the $g-r$ versus
$r$ data for all stars within $6\farcm0$ of the center.  The
background was estimated using stars located suitably far from the
object.  Figure~\ref{fig:jelte} shows the goodness-of-fit contours for
two different age bins, namely 4-5 and 8-10 Gyrs.  The inset shows the
improvement in the maximum-likelihood goodness-of-fit parameter Q
(Dolphin 2002) for the ages probed with respect to fitting a control
field CMD.  While the age of the stellar population is not
well-constrained, the best-fit metallicity and distance values do not
change significantly.  These fits imply a rather metal-poor stellar
population with [Fe/H]$\sim$-1.6 at a distance modulus of $\sim$23.3
magnitudes, corresponding to a heliocentric distance of $\sim 450$
kpc.  The fits are heavily weighted towards the most numerous
population in the CMD, in this case the RGB population.

\begin{figure}
\begin{center}
\includegraphics[width=0.35\textwidth,angle=-90]{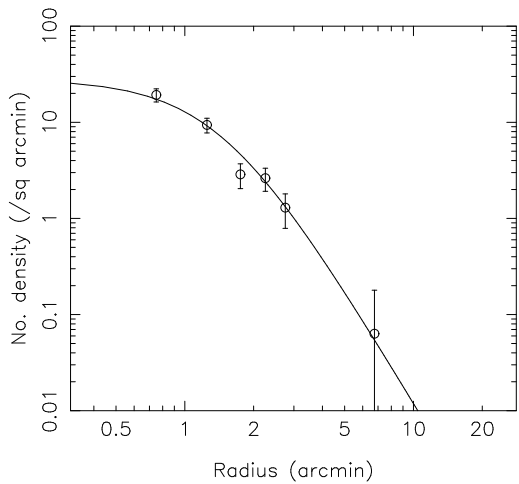}
\caption{Radial density profile of the \nameobj dwarf based on INT
data (all stellar objects with $g-r < 1.0$ and $g < 25.0$) together
with a Plummer law fit.
\label{fig:leo_prof}}
\end{center}
\end{figure}
\begin{figure}
\begin{center}
\includegraphics[width=0.4\textwidth,angle=-90]{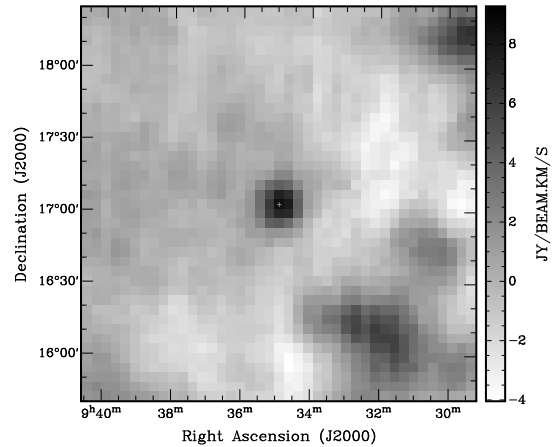}
\caption{Integrated HI flux density map centered on the coordinates of
the \nameobj dwarf. Significant HI is detected in two adjacent
velocity channels of data from the Northern HI Parkes All Sky Survey
(HIPASS, Wong et al. 2006). The above map is the sum of these two
adjacent velocity planes. The velocity channel separation of the
HIPASS data is 13.2 km s$^{-1}$. The integrated flux density within
the 15.5$^\prime$ beam centered on this source is 4.9 Jy km s$^{-1}$.
The detected point source, which has a flux-weighted heliocentric
velocity of 35 km s$^{-1}$, does not appear to be connected to the
undulating HI emission belonging to the Galactic interstellar medium.
\label{fig:leogas}}
\end{center}
\end{figure}

\section{Physical Properties and Stellar Population}

Follow-up observations of \nameobj were made on 30 November 2006 (UT)
using the 2.5 INT telescope with the WFC mosaic camera.  This contains
4 2k$\times$4k pixel EEV CCDs with a field of view of roughly 30$
\times$30 arcminutes and a scale of 0.33 arcseconds per pixel at the
field centre.  \nameobj was observed with single exposures of 1200s
and 900s in SDSS-like $g$, $r$ bands, and calibrated on the SDSS AB
system.  Data were processed in Cambridge using a general purpose
pipeline for processing wide-field optical CCD
data~\citep{Ir01}. Images were de-biased and trimmed, cross-talk
corrected, and then flatfielded and gain-corrected to a common
internal system using clipped median stacks of nightly twilight flats.
For each image frame an object catalogue was generated using the
object detection and parameterisation procedure discussed by Irwin et
al. (2004). Objects in the catalogues were morphologically classified
as stellar or non-stellar (or noise-like).  The detected objects in
each passband were then merged by positional coincidence (within 1
arcsec) to form a combined catalogue.
 
Figure~\ref{fig:leo_image} shows an INT image derived from the $g$ and
$r$ band data. The \nameobj dwarf is clearly visible, together with
its population of blue stars.  The left panels of
Figure~\ref{fig:leo_int} show the CMD within a $1\farcm8$ radius of
the centre. Even though the giant branch is sparsely populated, the
tip must be as bright as $i \sim 19.8$ (see Fig.~\ref{fig:leo_disc}),
which corresponds to $I \sim 19.3$, using the transformations given in
\citet{Sm02} and allowing an extra $0.06$ mag for extinction in the
i-band.  Using the calibration $I_{\rm TRGB} = -4.03$ \citep{Be02}, we
find that the distance modulus is therefore at most $23.3$.

The rightmost panel of Figure~\ref{fig:leo_int} shows the luminosity
functions of stars detected in the $g$-band image (see
Figure~\ref{fig:leo_image}), which is complete to $g \sim 25$.  The
black line shows the number of stars within a circle of radius
$4\farcm2$. This is compared with the grey line, which shows the
appropriately normalised luminosity function for the background
region.  There is an obvious excess of counts peaking at $g \sim
23.8$, which is where we would expect to see the horizontal branch.
This unambiguous detection, which is readily apparent in the CMDs
shown in the left panels, allows us to improve on the distance
estimate.  Assuming $M_{HB} = 0.6$ in the g-band at $g-r \approx$ 0
after allowing for extinction implies a distance modulus of
$23.1\pm0.1$, consistent with the sparsely populated RGB near the tip.
The middle left panel of Figure~\ref{fig:leo_int} shows overlaid
isochrones from \citet{Gi04}, corresponding to stellar populations of
age 200 and 500 Myrs at a distance modulus of 23.1.  Blue stars
brighter than $g=23$ and blueward of the giant branch are therefore
most likely to be helium burning blue loop stars associated with a
young component.  Two zones from the color-magnitude diagram,
indicated by the gray boxes were used for the spatial analysis.  The
RGB candidates were used to produce the density contours in the middle
right panel of the figure. The object is seen to be almost circular in
the inner parts.  The young blue stars are shown as filled circles in
this figure and are concentrated near the centre, thus confirming that
they are physically associated with the \nameobj dwarf.

To estimate the properties in Table~\ref{tab:struct}, we use the INT
data to derive the centroid from the density-weighted first moment of
the distribution and use the second moments to investigate the
ellipicity.  Athough circular in general appearance the ellipicity at
different contour levels averages $\sim 0.1$ but has a position angle
that varies by 90 degrees.  The radial profile shown in
Figure~\ref{fig:leo_prof} is derived from all stars with $g-r < 1$ and
satisfying $20 < g < 25$.  We compute the average density within
circular annuli after first subtracting a constant asymptotic
background level (0.8 arcminute$^{-2}$), reached at large radii, and
then fit the radial profile with a standard Plummer law.  At a
distance of $\sim 420$ kpc, the best-fitting half-light radius of
$1\farcm4$ corresponds to $\sim 170$ pc. This is the typical scale
length of some of the recently discovered Galactic dSph
galaxies~\citep{Be07}, but smaller than the classical Local Group dIrr
galaxies~\citep{vand00}.

The flux (AB magnitudes) in stars within a 5 arcmin radius relative to
a control background region is $g = 17.1$ and $r = 16.7$ computed by
integrating to the horizontal branch level. To estimate the
contribution from young stars between the horizontal branch and
main-sequence turn-off, we used the deep HST luminosity function of
Phoenix provided by \citet{Ho06} to obtain a correction of 0.5 mag. By
allowing 0.3 mag for stars fainter than turn-off, and 0.1 mag for
extinction, we obtain a total flux estimate of $g = 16.2$.  Using the
transformations in \citet{Sm02} and assuming the measured average
colour of $g-r = 0.4$ holds for the entire stellar population, this
equates to $V = 16.0$ (Vega).  At a distance modulus $(m-M)_0 = 23.1$,
this corresponds to an absolute magnitude of $M_V = -7.1$.  Assuming a
Plummer profile, this yields an observed $V$ band central surface
brightness of $\sim 27.0$ mag arcsec${}^{-2}$.  For the new dwarfs
this is a relatively high surface brightness, which is consistent with
the easy visibility on the image in Figure~\ref{fig:leo_image}

Finally, Figure~\ref{fig:leogas} shows an HI flux density map centered
on the coordinates of the \nameobj dwarf using data from the Northern
HI Parkes All Sky Survey (HIPASS, Wong et al. 2006). Located at the
same spot is a significant HI overdensity.  At a distance of $\sim
420$ kpc, the HI mass of this cloud would be $2\times10^5$
$M_\odot$. Its flux-weighted heliocentric velocity is 35 km s$^{-1}$,
corresponding to a Galactic Standard of Rest velocity of -61 km
s$^{-1}$, consistent with gentle Galactic infall.  It is clearly
isolated from the filamentary foreground of HI clouds, which have
similar velocities. The stellar center of \nameobj lies within the
same $4'\times4'$ HIPASS pixel that contains the centroid of the
HIPASS gas cloud.  This suggests that there is a high probability that
the cloud is physically associated with the dwarf. Although HIPASS
does not have adequate velocity resolution to measure the dispersion
directly the non-appearance of the compact HI cloud in adjacent
channels either side of the two detection channels constrains the
velocity dispersion to be less than $\sim 13$ kms$^{-1}$, which is
within the realm of low luminosity dwarfs.

The stellar luminosity of \nameobj is $\sim 4 \times 10^4 L_\odot$,
which corresponds to a mass of $\sim 10^5 M_\odot$, assuming a stellar
mass-to-light ratio of $\sim 2-3$. This gives a ratio of mass in HI to
mass in luminous material $M_{\rm HI} /M_L$ of about unity, consistent
with that found in dIrrs in the Local Group. The source is unresolved
in HIPASS, however the inferred average column density of HI within
the Plummer radius is $\sim 2 \times 10^{20} {\rm cm}^{-2}$. Although
this average is slightly below the threshold required for star
formation~\citep{El94}, the surface density of HI is likely to contain
local peaks~\citep{van97}.

\section{Conclusions}

We have discovered a new dwarf galaxy in the constellation of Leo,
which we have named \nameobj. Its CMD has a clear red giant branch,
from which we derive a tentative age of 6-8 Gyrs and a metallicity of
[Fe/H] $\sim -1.6$. Its distance, derived from the position of the
horizontal branch, is $\sim 420$ kpc.  Given this distance, \nameobj
is unlikely to have been strongly affected by the tides of the Milky
Way. This suggests that its very low luminosity is intrinsic and not
the result of disruption or stripping.  Clearly visible in the CMD is
a sequence of bright blue stars indicative of recent star
formation. Coincident with the position of the dwarf is an HI cloud of
$\sim 2 \times 10^5 \msun$. The \nameobj dwarf has some of the
characteristics of transition galaxies like Phoenix or Pisces, but,
with an absolute magnitude of $M_V \sim -7.1$, it is much fainter. It
is the least luminous galaxy found to date with recent
star-formation.

\acknowledgments Funding for the SDSS and SDSS-II has been provided by
the Alfred P.  Sloan Foundation, the Participating Institutions, the
National Science Foundation, the U.S. Department of Energy, the
National Aeronautics and Space Administration, the Japanese
Monbukagakusho, the Max Planck Society, and the Higher Education
Funding Council for England. The SDSS Web Site is
http://www.sdss.org/.  The SDSS is managed by the Astrophysical
Research Consortium for the Participating Institutions. The
Participating Institutions are the American Museum of Natural History,
Astrophysical Institute Potsdam, University of Basel, Cambridge
University, Case Western Reserve University, University of Chicago,
Drexel University, Fermilab, the Institute for Advanced Study, the
Japan Participation Group, Johns Hopkins University, the Joint
Institute for Nuclear Astrophysics, the Kavli Institute for Particle
Astrophysics and Cosmology, the Korean Scientist Group, the Chinese
Academy of Sciences (LAMOST), Los Alamos National Laboratory, the
Max-Planck-Institute for Astronomy (MPIA), the Max-Planck-Institute
for Astrophysics (MPA), New Mexico State University, Ohio State
University, University of Pittsburgh, University of Portsmouth,
Princeton University, the United States Naval Observatory, and the
University of Washington.

\end{document}